# New Architecture for Dynamic Spectrum Allocation in Cognitive Heterogeneous Network using Self Organizing Map

Himanshu Agrawal, and Krishna Asawa

Jaypee Institute of Information Technology, Noida, India
{himanshu.agrawal,krishna.asawa}@jiit.ac.in

**Abstract.** This paper introduces the Hybrid Architecture of Dynamic Spectrum Allocation in the hierarchical network combining centralized and distributed architecture to get optimum allocation of radio resources. It can limit the interference by interacting dynamically and enhance the spectrum efficiency while maintaining the desired QoS in the network. This paper presented dynamic framework for the interaction. The proposed architecture employed simple learning rule based on hebbian learning for sensing the primary network and allocating the spectrum.

## 1 Introduction

Electromagnetic radio spectrum is very precious natural resource. This air interface provides connectivity to millions of devices. With the introduction of latest handheld devices like smart phones, tablets; demand for spectrum has increased exponentially. Now it is common belief that soon there will not be any spectrum left for allocation. From the operators point of view, who has purchased mere 20 MHz band by spending millions of dollars wants to utilize every bit of the spectrum. But study [1] suggests that most of the licensed spectrum lies idle depending upon location and time. It reflects the limitations of current static allocation of frequency bands. It triggers the searching of new techniques for spectrum allocation that requires new regulation, policies, better economic environment. Cognitive radio [2] is such technique that provide dynamic spectrum access [3]; utilizing the unused spectrum fragments. Dynamic spectrum access is one of the most important applications of cognitive radio [4]. This study focuses on the hierarchical access model [5] which is a special class of dynamic spectrum access. In this model there is a consideration of secondary users (SU) or unlicensed users for whom the licensed spectrum is open to access, subject to the interference perceived by the licensed user or primary users (PU). There are two approaches of sharing the frequency bands: spectrum underlay or ultra wide band and spectrum overlay or opportunistic spectrum access [6]. In underlay approach the constraint on the transmission power of the SU so that it won't cause any interference to

PU. It is a very short range communication. It does not depend on the sensing of the surrounding environment. In overlay approach there is no restriction on the transmission power (theoretically) but SU can only transmit when the spectrum is not used by the PU. So it requires detection of the PU's signal. Opportunistic spectrum access consists of different steps involving spectrum sensing, sensed spectrum access and regulation policy. Spectrum overlay approach was proposed by Mitola [7]. This approach is compatible with the existing wireless technology and legacy systems. Moreover the spectrum overlay and underlay networks can coexist and further improve the spectrum utilization.

The organization of this paper is as follows. Dynamic spectrum allocation is discussed in section 2, section 3 describe the proposed architecture, section 4 describe the problem definition, system model and mathematical formulation, section 5 describe existing techniques of spectrum allocation, proposed approach of spectrum allocation using self-organizing feature map, it's architecture and algorithm and in section 6 this paper is concluded and some future aspects are also discussed.

## 2    Dynamic Spectrum Allocation

Dynamic Allocation of Spectrum to a secondary user (SU) must prevent the interference caused to the primary user (PU). The dynamic allocation of spectrum is to assign the required bandwidth to each SU in such a way that interference is minimized and spectrum utilization is maximized. It can be achieved by opportunity identification [8] that provide the information about the available spectrum bands.

### 2.1    Opportunity Identification

It is very essential for dynamic spectrum allocation (DSA). A fragment of spectrum can be considered as an opportunity, if it is not occupied by primary users currently. There are different constraints to be satisfied in order to identify any channel as opportunity. Suppose there are two secondary users A and B, situated geographically far apart as per Figure 1 [9]. They are surrounded by different primary users. If they are able to communicate over a channel without causing any harmful interference to nearby primary users then this channel will be considered as an opportunity. User A wishes to transmit to user B then A has to take care of nearby primary receivers while B has to protect nearby primary transmitters. In this case there should not be any primary user who is transmitting within a range of $r_{tx}$ from A and receiving within a range of $r_{rx}$ from B. Obviously $r_{tx}$ is related with transmission power of A and depend on the interference toleration limit of primary users and $r_{rx}$ is determined by transmission power of primary users and secondary user's interference toleration limit. Interference toleration limit is part of spectrum regulatory policy. There are certain important points like spectrum opportunity depend upon the geographical location of secondary users and primary users.

### 2.2    Opportunity Sharing

Opportunity or Spectrum sharing among secondary users requires coordination else control by centralized entity. So the architecture can be classified as centralized or distributed. In centralized architecture all the procedure of spectrum allocation is done by centralized entity while spectrum sensing is distributed and perform by associated secondary users [10, 11]. It requires cognitive capabilities in the secondary users while in proposed architecture spectrum sensing and allocation is performed by Spectrum Controller (SC). There are several spectrum controllers in the network. It is hybrid architecture. It is more compatible with the legacy systems than any other architecture because it does not requires cognitive capabilities in the secondary users. In distributed architecture [12] sensing and allocation is performed by individual node distributively. It requires lots of message exchange between nodes which is an overhead. In the proposed architecture it requires less amount of message exchange because it takes place between spectrum controllers only. There is another classification based on user's behavior i.e. cooperative or non-cooperative. In cooperative spectrum sharing, nodes share their respective sensing information with other nodes and improve the sensing results by solving the problem of hidden terminals [13]. In the proposed architecture spectrum controllers cooperate with each other and improve the overall sensing results. In non-cooperative spectrum sharing, interference at other nodes is not considered. It provides selfish solutions [8, 13]. Cooperative approach provide better solutions but with more number of message exchange takes place.

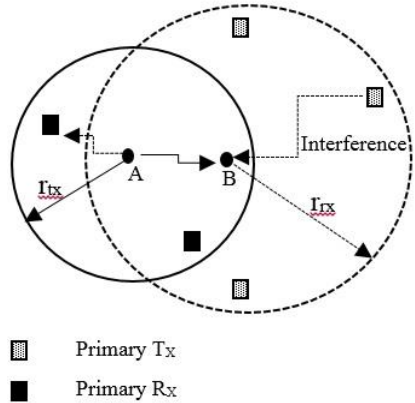

**Fig. 1.** Definition of Spectrum Opportunity where A and B are secondary users, A is tx B is rx.

## 3      Proposed Architecture

The availability of number of channels may vary with respect to time due to dynamic nature of wireless network and primary user may appear at any point of time. So the spectrum allocation schemes must be adaptive to these frequent changes. This study considers the heterogeneous network consisting two types of nodes first is spectrum controller and second is unlicensed user as per Figure 2. In this architecture the spectrum controller exploit the opportunities in primary network's unused bands and then allocates them to unlicensed user. It is responsible for spectrum opportunity identification and then it allocates it to SUs to achieve the objectives like efficiency, fairness and maintain the status of no interference to primary user.

Opportunity identification is done by cooperation among spectrum controllers (SC). It will improve the quality of spectrum sensing and at the same time probability of collision with PU also minimized. However it affects the spectrum efficiency because spectrum needed for cooperation traffic. In particular SC will detect the PU signal by using feature detection or *cyclostationarity* method [14]. It is done by analyzing spectral correlation function. It is robust method and able to deal with random noise power. However it is very complex and requires past observations and sufficient computational resources. The spectrum controller (SC) provides the opportunistic spectrum access to its associated secondary users. When SC senses transmission free channel it estimate its capacity and its valuation. Spectrum controllers compete for limited number of resources with each other i.e. self-coexistence.

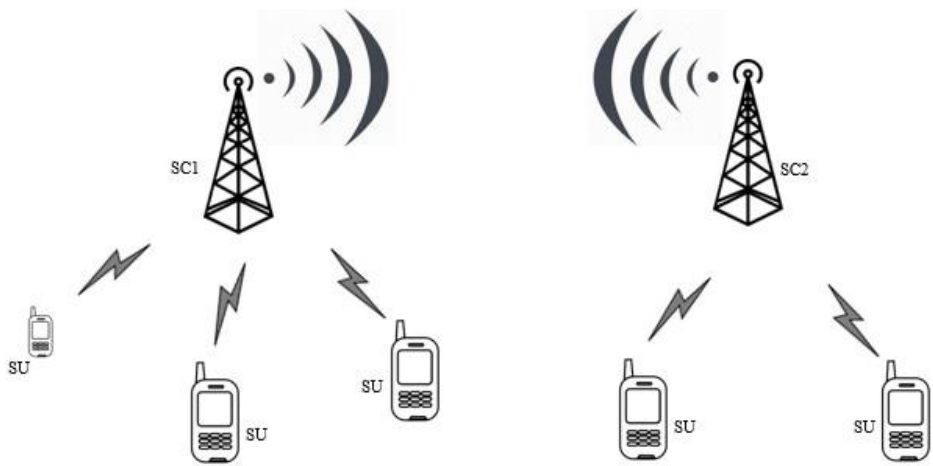

**Fig. 2.** The System Model for Spectrum Sharing among Secondary Users

### 3.1 Cooperative Sensing

The accuracy of spectrum sensing is foremost important and it must be reliable. Cooperative sensing [8, 13] is proposed in the literature that can solve the problem of hidden node, shadowing and propagation delay. It uses the spatial diversity to detect the PU activities. In this method cognitive radio users exchange the sensing information and by combining them they have more reliable information about PU activities. There are two methods of cooperative sensing, distributed and centralized. There are respective pros and cons of each method like centralized need backbone infrastructure that limit the scalability of network while in distributed approach lot of signaling takes place between cognitive radio units that affect spectrum efficiency. This study proposes to use hybrid approach in which different spectrum controllers cooperate with each other and decide the available opportunities as per Figure 3. The objective of this method is to ensure the protection of PU and exploit the available opportunities as well.

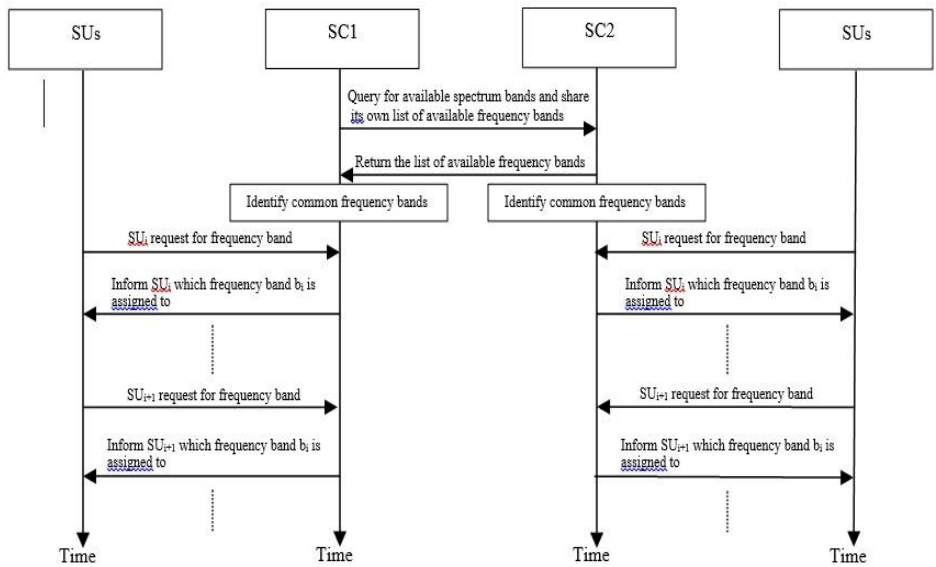

**Fig. 3.** Illustration of the operation of the Cognitive Network

The proposed architecture requires dedicated communication link between different spectrum controllers they may communicate over unlicensed band or opportunistic spectrum band. It is an overhead to provide communication link between different spectrum controllers to exchange sensing information. It can be referred as common control channel. The advantage of this approach is; SC can act as spectrum broker that can collect the revenue. It is a robust approach to exploit the opportunities in time domain. The proposed approach can achieve the desired quality of service as well by implementing priority scheduling of spectrum allocation based on revenue provided by the secondary user. This architecture achieves fairness in allocation of spectrum and minimizes the energy consumption of end user at the same time by saving energy in spectrum sensing. It is a cheat proof spectrum sharing scheme.

It is proposed in this paper to use store and forward approach for data transfer i.e. data from the user may be stored at SC and at the most suitable time when spectrum is free like at midnight it may be forwarded. This study proposed this architecture because the cognitive capability at each and every user is very costly so if this capability can be shifted to a centralized controller then the end user can be considered as an ordinary mobile or fix terminal and the required cost will also reduce.

This proposal can efficiently address the following challenges: (i) interference avoidance – secondary users should access the licensed spectrum without interfering the primary users. (ii) QOS awareness – SC can select and allocate appropriate spectrum band for communication depending upon the desired QOS requirement. (iii) Seamless communication – SC can provide the seamless communication even if the PU appears, by switching to some other vacant frequency band i.e. spectrum mobility can be managed by the SC.

## 4   Problem Definition

This study considers noninterference constraint as soft and demand satisfaction constraint as hard in the objective function. It is useful to identify solutions which are not interference free but at least exist in situation where no conflicts free assignment possible.

The spectrum allocation problem (SAP) can be formulated as

(SAP)    minimize        degree of interference
         subject to      spectrum requirement constraints

SAP = Minimize interference subject to maximize utilization of spectrum

So, the solutions should limit the interference to the PU below certain threshold and at the same time allocate the spectrum to different SU fairly and efficiently. In real life scenario the number of available channels is very less than the actual requirement. So the solution should be able to accommodate as many demands as possible.

### 4.1   System Model

A geographical area is considered that consist of $S$ spectrum controllers (SC) and $C$ available channels. A pool of spectrum bands is considered at every individual spectrum controller. Each band is of equal bandwidth. It is assumed that primary network is a cellular network. In the secondary network the spectrum controller nodes are cooperative and share their sensing data. This cooperative sensing can solve the problem of hidden nodes and improve the results as well. The SC nodes may communicate over the unlicensed bands to avoid interruption by the PUs. They are in continuous touch to share the sensing information. Interference is major constraint that needs to be taken care. The interference constraints can be described by $I$ matrix called interference matrix i.e. $I = \{i_{n,k,m} | i_{n,k,m} \in \{0,1\}\}_{SXSXC}$. It is $S$ by $S$ by $C$ matrix representing the interference, if two users use the same channel simultaneously i.e. $i_{n,k,m} = 1$.

Requirement for spectrum at each spectrum controller is represented by $R$. It is an $S$ dimension vector. Dynamic network conditions are modeled by time varying entries in $C$ and $R$.

### 4.2 Mathematical Formulation

The channel assignment matrix can be defined as

$$A_{n,m} = \begin{cases} 1, & \text{if SC } n \text{ is assigned to channel } m \\ 0, & \text{otherwise} \end{cases}$$

for $n = 1, \ldots, S$ and $m = 1, \ldots, C$. Amount of interference can be measured by assigning a weight to each assignment. The proximity factor $P$ can be defined as $P_{n,k,d+1}$ where $n, k$ are spectrum controllers and $d$ is distance between them. If it is zero then the interference cost would be high. So it can be formulated to minimize the overall cost of the network.

minimize

$$f(x) = \sum_{n=1}^{S}\sum_{m=1}^{C} A_{n,m} \sum_{k=1}^{S}\sum_{j=1}^{C} P_{n,k,d+1} A_{k,j}$$

subject to

$$\sum_{m=1}^{C} A_{n,m} = B_n, \quad \forall n = 1,\ldots,S$$

$$A_{n,m} \in \{0,1\} \quad \forall n = 1,\ldots,S, \text{ and}$$

$$m = 1,\ldots,C.$$

The cost function can be generated as

$$P_{n,k,d+1} = \max(0, P_{n,k,d} - 1)$$

$$P_{n,k,1} = I_{n,k,m}, \quad \forall n, n \neq k, \text{ and}$$

$$m = 1,\ldots,C.$$

$$P_{n,n,1} = 0, \quad \forall n.$$

## 5  Spectrum Allocatıon usıng Self Organızıng Feature Map

It has been shown in the literature that spectrum allocation is NP-hard problem [6] and may be mapped to a graph coloring problem [15]. Many techniques proposed in the literature to solve this problem. The techniques can be classified in terms of heuristics based, graph theory based, game theory based, linear programming based, fuzzy logic based and evolutionary algorithm based. The limitations of heuristics based technique [16] are that they are problem specific and might converge to local minima, although they are easy to implement and simple. Graph theory based algorithms cannot include all parameters like QoS but they can use existing techniques to solve the problem like graph coloring. In game theory [10], it is difficult to reach to equilibrium although it can model cooperation and non-cooperation with high accuracy. Linear programming techniques [17] requires many assumptions but able to use existing solutions. Fuzzy logic based techniques [18] can provide quick decisions but requires lots of rules and not feasible to incorporate the dynamics of the network. The techniques based on evolutionary algorithms [19] are slow and may stuck in local minima although they are able to hold different constraints and objectives.

### 5.1  Self Organızıng Feature Map

It is a special class of neural network. It requires less number of neurons and weights. It is inspired by human brain. It exploits the principal of ordering of information like brain's ability to find the structure in the information. In a self-organizing feature map (SOM) [20] neurons are selectively tuned to the various input patterns. They compete with each other to claim the input and weights are adapted to reflect the pattern in the input space. There are three fundamental elements of SOM: self-amplification, competition and inhibition. Basically learning in self-organizing system is based on these concepts. In SOM the neurons process the information in parallel and they are able to find structure and order from information.

This paper proposes a new self-organizing neural network based dynamic spectrum allocation architecture. The basic concept behind this proposal is; in the hierarchical network, nodes are competing to assign the channels. So this competition is realized through a self-organizing map.

### 5.2  SOFM approach to the SAP

The architecture of Self Organizing Feature Map is shown in Figure 4. There are $S$ nodes in the input layer that represent the number of spectrum controllers (SC) in the network and there are $C$ nodes in the output layer that represent the number of channels in the network. The weight $W_{i,j}$ is assigned to connection of node $i$ of output layer to node $j$ of input layer. Basically weight vector $W_{i,j}$ represent the probability that the value of an assignment in $i^{th}$ row and $j^{th}$ column in matrix $A_{i,j}$ is 1. It is probability matrix $W$ on which the Self Organizing Feature Map by Kohonen [21] is applied. The spectrum requirement for each of the spectrum controller is presented to the network after that the output nodes are competing with each other to claim that input. After that weight adaptation takes place and neighbourhood of the winning neuron is identified. For example if single channel requires at spectrum controller $j'$ then the value of input vector is zero for all input neurons except for which demand is raised, input will be "one" for that spectrum controller. Now the value of objective function is calculated for each of the neurons of output layer. The objective function can be defined as –

$$Y_{i,j'} = \sum_{k=1}^{S}\sum_{j=1}^{C} P_{j',k,d+1} W_{k,j}$$

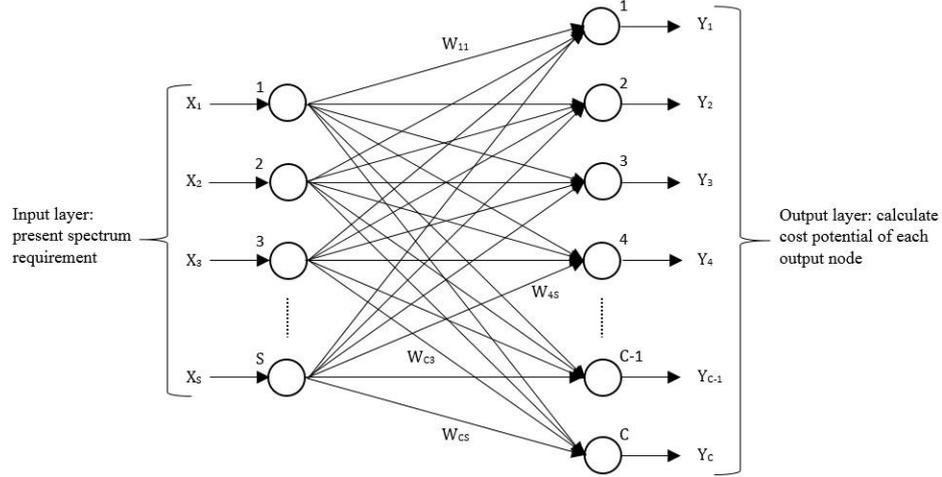

**Fig. 4.** Architecture of Dynamic Spectrum allocation using Self Organizing Feature Map

The neuron for which the value of objective function is minimum will be the winning neuron $c_0$,

$$Y_{c_0,j'} \leq Y_{i,j'} \quad \forall i \text{ and } j'$$

The neighbourhood of the winning neuron $c_0$ is defined as the set of nodes for which the value of objective function is in increasing order. Size of the neighbourhood is denoted by $\eta_{j'}$ for node $j'$ i.e.

$$Y_{c_0,j'} \leq Y_{c_1,j'} \leq Y_{c_2,j'} \leq Y_{c_3,j'} \leq Y_{c_4,j'} \leq \ldots \ldots \leq Y_{c_{\eta_{j'}},j'}$$

Now the weights are adapted according to the Kohonen's rule and it will be within the neighbourhood of winning neuron. Size of neighbourhood depend on the input presented. Initially its size is large but gradually decreases with the update of the weights. It decreases until its value is equal to demand of spectrum i.e. $\eta_j = R_j \quad \forall j$. To prevent the violation of constraints by the weight matrix, an approach of random update of weights is followed such that the following energy function will be minimized.

$$\varepsilon = ||w - (Pw + s)||^2$$

Where

$$P = I - A^T(AA^T)^{-1}A$$

$$s = A^T(AA^T)^{-1}b$$

The constraint plane is defined as

$$Aw = b$$

Where

$$A = \begin{bmatrix} \overbrace{1\ 1\ \ldots\ 1}^{C} & \overbrace{0\ 0\ \ldots\ 0}^{C} & \cdots & \overbrace{0\ 0\ \ldots\ 0}^{C} \\ \overbrace{0\ 0\ \ldots\ 0}^{C} & \overbrace{1\ 1\ \ldots\ 1}^{C} & \cdots & \overbrace{0\ 0\ \ldots\ 0}^{C} \\ \vdots & \vdots & & \vdots \\ \overbrace{0\ 0\ \ldots\ 0}^{C} & \overbrace{0\ 0\ \ldots\ 0}^{C} & \cdots & \overbrace{1\ 1\ \ldots\ 1}^{C} \end{bmatrix}$$

In above matrix there are $S$ rows. $b$ is demand of spectrum in the network while $w$ is weight. Energy function should be minimized such that the $w$ lies on the constraint plane.

### 5.3    Algorıthm of Spectrum Allocatıon usıng Self Organızıng Feature Map

1. Choose Initial weights of SOM according to following condition.
$$W_{j,k} = \frac{R_j}{C}$$
2. Randomly generate initial spectrum requirements and represent it as input vector *x*.
3. Now calculate the objective function for each of the node of output layer.
4. Identify the winning node and its neighbourhood.
5. Update the weights of neighbouring node of winning node as follows –

$$W'_{j,k} \leftarrow W_{j,k} + \Delta W_{j,k}$$

Where

$$\Delta W_{j',k} = \alpha(\eta, t)[1 - W_{j',k}] \quad \forall k: Y_{k,j'} < Y_{c_{\eta_{j'}}, j'}$$

Where

$$\alpha(\eta, t) = \frac{\alpha(t)\rho_{j'}}{R_{j'}} exp\left[-\frac{|Y_{c_0,j'} - Y_{k,j'}|}{\sigma(t)}\right]$$

Where $\alpha$ and $\sigma$ are functions of time and $\rho$ is weight vector. Nodes which are not included in neighbourhood, $\Delta W_{j,k} = 0$.

6. Weights must lie on constraint plane. Weight are adapted such that $Aw = b$.
7. Goto step 2 again and take another spectrum requirement. Repeat this loop till all the requirements for all the spectrum controllers are presented to the network. This is one iteration. Repeat it for $N$ iterations.
8. Repeat step 7 till there is no change in weight takes place i.e. $\Delta W_{j,k} \approx 0$.
9. Repeat all above steps till neighbourhood size is equal to demand of spectrum i.e.

$$\eta_j = R_j \quad \forall j$$

Definition of parameters – Weight vector $\rho$ is defined as follows

$$\rho_i = \left(\sum_{j=1}^{S} R_i I_{ij}\right) - I_{ii}, \quad \forall i = 1, \ldots, S$$

Remaining parameters $\alpha$ and $\sigma$ are defined as

$$\alpha(0) = \min R_i$$
$$\alpha(t+1) = 0.95\alpha(t)$$
$$\sigma(0) = 9$$
$$\sigma(t+1) = 0.95\sigma(t)$$
$$\eta_j(0) = R_j + int\frac{S}{5}$$
$$\eta_j(t+1) = \eta_j(t) - 1$$

# 6    Conclusion

This paper considers the assignment of spectrum to unlicensed users and at the same time prevent the harmful interference to primary users. It achieves optimal efficiency in allocation of spectrum. Interference constraints and demand of spectrum is taken into consideration in the objective function. This study propose to use modified version of Kohonen's Self Organizing Feature Map in order to optimize Spectrum Allocation in the network. This paper shows the significance of neural network approach to solve optimization problem in real life scenarios. Due to their inherent adaptive nature these approaches are particularly very useful in adaptive operations in wireless networks. This approach can also handle the incoming requests of variable bandwidth requirements. It will be the subject of future research.